\documentclass{article}
\usepackage{amsmath}
\usepackage{algorithm}
\usepackage{algpseudocode}
\usepackage{amssymb}
\usepackage[most]{tcolorbox}
\usepackage{xcolor}
\usepackage{adjustbox}
\usepackage{booktabs}
\usepackage{array}
\usepackage{wrapfig}
\usepackage{enumitem}
\usepackage{fancyvrb}
\usepackage[table]{xcolor}
\definecolor{lightblue}{RGB}{230,245,255}
\definecolor{lightred}{RGB}{255,230,230}
\definecolor{darkgreen}{RGB}{0,100,0}
\tcbuselibrary{listingsutf8}

\usepackage{listings}
\usepackage{xcolor}

\usepackage{upquote}

\definecolor{codebg}{RGB}{248,248,248}
\definecolor{codeframe}{RGB}{220,220,220}
\definecolor{codekw}{RGB}{0,92,197}
\definecolor{codestr}{RGB}{196,26,22}
\definecolor{codecom}{RGB}{0,128,0}
\definecolor{codenum}{RGB}{120,120,120}

\lstdefinestyle{covenant}{
  backgroundcolor=\color{codebg},
  frame=none,
  basicstyle=\ttfamily\small,
  keywordstyle=\color{codekw}\bfseries,
  stringstyle=\color{black}, 
  commentstyle=\color{codecom}\itshape,
  showstringspaces=false,
  showspaces=false,
  showtabs=false,
  keepspaces=true,
  upquote=true,
  columns=fullflexible,
  breaklines=true,
  breakatwhitespace=false,
  breakindent=0pt,
  tabsize=2,
  xleftmargin=1mm,
  xrightmargin=1mm,
  aboveskip=0.5em,
  belowskip=0.5em,
  postbreak=\mbox{\textcolor{gray}{$\hookrightarrow$}\space}
}
\lstdefinestyle{covenantpython}{
  style=covenant,
  language=Python,
  morecomment=[s]{"""}{"""},
  morecomment=[s]{'''}{'''}
}
\newtcolorbox[auto counter, number within=section]{functionbox}[2][]{colback=lightblue!10!white, colframe=blue!75!black, fonttitle=\bfseries, title=Function: #2,#1}
\newtcolorbox[auto counter, number within=section]{functionbox2}[2][]{colback=lightgreen!10!white, colframe=green!75!black, fonttitle=\bfseries, title=Function: #2,#1}

\definecolor{todoRed}{HTML}{8F2D2D}
\definecolor{covenantred}{HTML}{FF3A3A}
\definecolor{covenantgreen}{HTML}{2ECC71}
\definecolor{covenantblk}{HTML}{101010}
\definecolor{covenantblk800}{HTML}{2F2F2F}
\definecolor{covenantgray2}{HTML}{828282}
\definecolor{covenantwhite}{HTML}{F4F4F4}
\definecolor{covenantgrid}{HTML}{DDDDDD}

\tcbset{enhanced}

\definecolor{covenantgray}{gray}{0.92}
\newcolumntype{G}{>{\columncolor{covenantgray}}c}

\usepackage[preprint]{neurips_2024}

\usepackage[utf8]{inputenc} 
\usepackage[T1]{fontenc}   
\usepackage{hyperref}      
\usepackage{url}            
\usepackage{booktabs}     
\usepackage{amsfonts}       
\usepackage{nicefrac}      
\usepackage{microtype}      
\usepackage{xcolor}         
\usepackage{pgfplots}       
\pgfplotsset{compat=1.18}
\usepgfplotslibrary{groupplots}

\title{Covenant-72B: Pre-Training a 72B LLM with Trustless Peers Over-the-Internet }

\author{
\textbf{Joel Lidin}$^{1}$\thanks{Correspondence to \texttt{joel@tplr.ai}} \quad
\textbf{Amir Sarfi}$^{1}$ \quad
\textbf{Erfan Miahi}$^{1}$ \quad
\textbf{Quentin Anthony} \quad
\textbf{Shivam Chauhan}$^{1}$ \quad\\
\textbf{Evangelos Pappas}$^{1}$ \quad
\textbf{Benjamin Th\'erien}$^{2}$ \quad
\textbf{Eugene Belilovsky}$^{2}$ \quad
\textbf{Samuel Dare}$^{1}$ \\
\\[0.5em]
$^{1}$Covenant AI \quad
$^{2}$Mila
}
\begin{document}

\maketitle

\begin{abstract}
    Recently, there has been increased interest in globally distributed training,
    which has the promise to both reduce training costs and democratize
    participation in building large-scale foundation models. However, existing
    models trained in a globally distributed manner are relatively small in scale
    and have only been trained with whitelisted participants. Therefore, they do
    not yet realize the full promise of democratized participation. In this
    report, we describe \textsc{Covenant-72B}, an LLM produced by the largest
    collaborative globally distributed pre-training run (in terms of both compute
    and model scale), which simultaneously allowed open, permissionless
    participation supported by a live blockchain protocol. We utilized a
    state-of-the-art communication-efficient optimizer, SparseLoCo, supporting dynamic participation with peers joining and leaving freely. Our
    model, pre-trained on approximately 1.1T tokens, performs competitively with
    fully centralized models pre-trained on similar or higher compute budgets,
    demonstrating that fully democratized, non-whitelisted participation is not
    only feasible, but can be achieved at unprecedented scale for a globally
    distributed pre-training run.
\end{abstract}

\section{Introduction}
\label{sec:intro}
LLMs have achieved remarkable success through scaling the model and dataset sizes~\cite{brown2020language,chowdhery2023palm,hoffmann2022training}, leading to models being trained at unprecedented scales and significant expense~\cite{grattafiori2024llama3herdmodels,liu2024deepseek,kimiteam2025kimik2openagentic}. Yet, state-of-the-art pre-training still largely depends on synchronous optimization with communication at every step, which drives the costly, tightly-coupled hardware buildouts (e.g., thousands of accelerators connected by high-bandwidth, low-latency interconnects). As a result, pushing to the largest scales remains dominated by organizations able to finance and operate highly centralized infrastructure. \emph{Decentralized} training makes large-scale pre-training accessible to geographically distributed compute connected via commodity internet links. This allows many lower-cost resources, potentially contributed by diverse actors, to be pooled together, which in turn lowers the barrier to entry and democratizes participation in training large-scale models.

However, achieving high training efficiency in this setting presents a significant challenge due to bandwidth constraints, higher latency, and dynamic participation. Existing efforts demonstrate that over-the-internet training can work in this challenging regime, but they fall short of fully democratizing participation because they rely on whitelisted contributors~\cite{dedloc,intellect1}. A promising step toward removing this restriction was presented in~\cite{gauntlet}, which demonstrated training with open participation from untrusted compute providers. Ultimately, making large-scale training communication-efficient under \emph{permissionless} participation is key to truly democratizing foundation model development and reducing dependence on centralized infrastructure.

In this report, we present \textsc{Covenant-72B}, the largest collaborative decentralized pre-training run to date and one of the first to run on a trustless compute network. \textsc{Covenant-72B} demonstrates successful large-scale LLM pre-training by infrequently communicating heavily compressed (more than $146\times$) pseudo-gradients. The run relies on two crucial components: \emph{Gauntlet}~\cite{gauntlet} and SparseLoCo~\cite{sparseloco}. Gauntlet is a mechanism for rewarding peers for contributing compute to the run and incentivizing honest participation. SparseLoCo is a recently introduced communication-efficient optimizer, known for its Pareto-optimal performance-communication tradeoff. The final model was trained for ${\sim}1.1$T tokens and achieves results competitive with centralized models trained at a similar scale, despite being trained over standard internet links and with dynamic participation (peers may join and leave freely). To further evaluate the model's capabilities, we perform a short ${\sim}14.8$B-token Supervised Fine-Tuning (SFT) stage, leading to performance competitive with similar chat models. We open-source intermediate and final pre-training checkpoints and post-training checkpoints under an Apache License.

In what follows, Section~\ref{sec:methodology} describes the optimization and incentivization methodology; Section~\ref{sec:systems} presents the communication protocol and systems design; and Sections~\ref{sec:training_details} and~\ref{sec:sft} evaluate the performance of our base and SFT models.

\section{Background and Methodology}
\label{sec:methodology}

\begin{figure}[t]
    \centering
    \includegraphics[width=\linewidth]{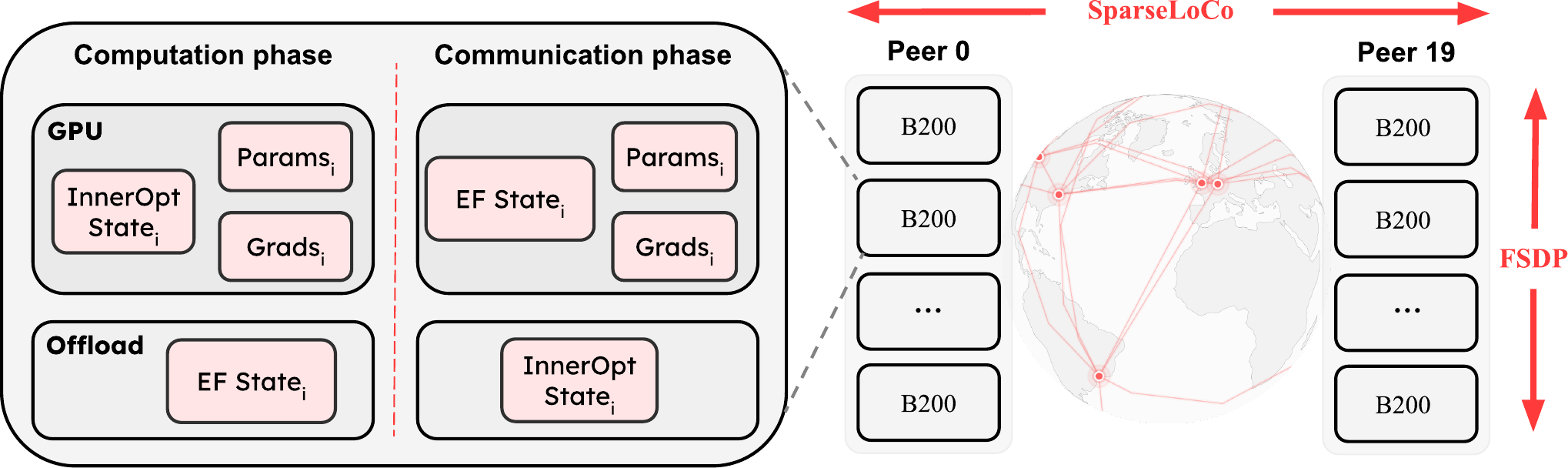}
    \caption{\textbf{\textsc{Covenant-72B} parallelism protocol.} Each peer runs
        a SparseLoCo replica and communicates heavily compressed and
        2-bit-quantized pseudo-gradients with other peers. Within each peer,
        $8\times$B200 GPUs use dynamic FSDP to shard model parameters, gradients,
        and training states across local GPUs. During the computation phase (inner
        steps), GPU $i$ requires only the inner optimizer state shards
        \texttt{InnerOpt State$_i$} while the error-feedback \texttt{EF State$_i$}
        is offloaded. During the communication phase, \texttt{InnerOpt State$_i$} is
        offloaded and swapped with \texttt{EF State$_i$} to compute compressed
        pseudo-gradients and update the error-feedback buffer. }
    \label{fig:main}
\end{figure}

\textsc{Covenant-72B} combines a communication-efficient optimizer (SparseLoCo) with a permissionless incentivization mechanism (Gauntlet). This section summarizes both methods and the adaptations required at 72B scale.

\subsection{SparseLoCo}
\label{sec:sparseloco_background}

SparseLoCo~\cite{sparseloco} is a local-update distributed optimizer that uses Top-$k$ sparsification, error-feedback, and quantization to reduce communication while outperforming dense baselines (e.g., DiLoCo~\cite{diloco,scalingdiloco}).
Let $\mathcal{R}=\{1,\dots,R\}$ denote the set of participating workers and let $\theta^{(t)}$ denote the global parameters at outer round $t$.
At each round, worker $r\in\mathcal{R}$ offloads a copy of the current synchronized model parameter state $\theta_r^{(t,0)}$ and runs $H$ steps of an inner optimizer (e.g., AdamW) on its assigned data shard $\mathcal{D}_r$, producing local models with parameters $\theta_r^{(t,H)}$. It then computes pseudo-gradients and compresses them using Top-$k$ sparsification, quantization, and error-feedback:
\begin{equation}
    \label{eq:inner}
    \begin{aligned}
        \Delta_r^{(t)}           & = \theta^{(t)} - \theta_r^{(t,H)}                          \\
        \widehat{\Delta}_r^{(t)} & = Q(\textsc{Top-}k(\beta\,e_r^{(t)}+\Delta_r^{(t)}))       \\
        e_r^{(t+1)}              & = \beta\,e_r^{(t)}+\Delta_r^{(t)}-\widehat{\Delta}_r^{(t)} \\
    \end{aligned}
\end{equation}
Here, $e_r^{(t)}$ is the local error-feedback state, $\beta$ is the error-feedback decay, and $Q(\cdot)$ is a low-bit quantizer. Intuitively, the error-feedback buffer accumulates the portion of $\Delta_r^{(t)}$ not transmitted in the current round, enabling aggressive sparsification without information loss.

All replicas then aggregate the compressed pseudo-gradients $\widehat{\Delta}_r^{(t)}$ from other peers and advance their local models, leading to the same global model $\theta^{(t+1)}$ on all replicas:
\begin{equation}
    \label{eq:outer}
    \begin{aligned}
        \Delta^{(t)}   & = \frac{1}{R}\sum_{r\in\mathcal{R}} \big(\widehat{\Delta}_r^{(t)}\big), \\
        \theta^{(t+1)} & = \theta^{(t)} - \alpha\,\Delta^{(t)} .
    \end{aligned}
\end{equation}

In practice, Top-$k$ sparsification is applied independently per tensor or layer~\cite{lin2017deep,sahu2021rethinking}. Doing so can create substantial index overhead and complicate integration with standard model parallelism methods. SparseLoCo instead uses a \emph{chunk-wise} Top-$k$ operator: each 2D tensor is partitioned into non-overlapping $64\times 64$ blocks and each 1D tensor into contiguous chunks of size $4096$, and Top-$k$ is applied separately within each chunk. Chunking has two practical benefits: (i) it aligns naturally with standard parallelism methods such as tensor parallelism (TP) and fully sharded data parallelism (FSDP), since compression can be performed independently on each shard, and (ii) it reduces index-transmission overhead by shrinking the local index space.

Since Top-$k$ selection is limited to chunks with $C$ unique positions, the information-theoretic lower bound for encoding the selected indices is
$\frac{1}{k}\log_2 \binom{C}{k}$ bits per transmitted value (equivalently $\log_2 \binom{C}{k}$ bits per chunk).
For our configuration ($C{=}4096$, $k{=}64$), this bound is approximately $7.36$ bits/value. In practice, we found that implementing an encoding scheme that approaches this degree of encoding introduces significant overhead. Thus, the chunk size being small becomes advantageous as it allows us to reduce the total index overhead to $12$ bits/value without a complex encoding scheme.

\subsection{Gauntlet}
\label{sec:gauntlet_background}
Gauntlet~\cite{gauntlet} enables permissionless training coordinated using a blockchain protocol by introducing a validator that scores submitted pseudo-gradients and selects which participants contribute to the global aggregation each round and broadcasts them to the network. The main evaluation signal, $\texttt{LossScore}$, comes from forwarding small batches of data and computing the loss difference before and after applying each participant's contribution. This is made efficient by evaluating only a subset of peers on a small subset of data, and by maintaining a persistent OpenSkill~\cite{joshy2024openskill} ranking over time to stabilize scores under per-round randomness. In parallel, the validator runs \emph{fast} checks on all participants based on their pseudo-gradients (e.g., liveness, synchronization with the main model, etc.), and combines these signals into a final score used to select the round’s contributors. Each peer on the network is assigned a (potentially overlapping) subset of data. To ensure peers are training on their assigned data, the submitted pseudo-gradients are further evaluated by comparing their loss improvement ($\texttt{LossScore}$) on a small subset of assigned and unassigned data. If they improve the loss on random data more than on their assigned data, the peer receives a negative score. This helps prevent participants from copying others or submitting duplicate behavior.

The system is calibrated such that there are always slightly more active participants than aggregated contributors so that any peer that drops out is quickly replaced. This is illustrated in Appendix~\ref{app:participation}.
To ensure additional robustness to peers with errors or adversarial behavior in an open-participation setting, we additionally normalize the pseudo-gradient of individual submissions before aggregation. Pseudo-gradient contributions are scaled relative to their \emph{median} norm so that no single participant can dominate the aggregation due to an abnormally large-magnitude update.

\section{Communication Protocol and Systems}
\label{sec:systems}

\paragraph{Hardware and parallelism.}
In \textsc{Covenant-72B}, each peer runs a \emph{SparseLoCo replica} and the
cross-peer communications occur through SparseLoCo's heavily compressed
pseudo-gradients. In this system, peers were expected to have at least $8\times$
NVIDIA B200 GPUs, with some participants contributing larger compute. Within
each peer, we use dynamic Fully Sharded Data Parallel (FSDP) across all local
GPUs to shard model parameters, gradients, and the inner optimizer state.
SparseLoCo introduces an additional outer error-feedback buffer
(Eq.~\ref{eq:inner}), and we shard the outer error-feedback buffer across GPUs
using the same FSDP partitioning strategy used for the inner optimizer states.

During the computation phase, we only require the inner optimizer states to
execute the $H$ inner steps, while the error-feedback buffer can be offloaded.
During the communication phase, inner optimizer states are no longer needed and
can be offloaded, while the sharded error-feedback buffer is swapped onto the
GPU to compute compressed pseudo-gradients and be updated (Eq.~\ref{eq:inner}).
We dynamically manage these phase-dependent optimizer state offloading and
ensure unused parameters and states do not remain resident on GPU and consume
memory when not needed.

Within the communication phase, after compressed pseudo-gradients are calculated
and the error-feedback is updated (Eq.~\ref{eq:inner}), the error-feedback is no
longer needed for the actual model update (Eq.~\ref{eq:outer}). Therefore, while
the compressed pseudo-gradients are communicated, the inner optimizer states are
swapped back with the error-feedback state to overlap the transfer with
communication. The parallelism strategy is summarized in Figure~\ref{fig:main}.

\paragraph{Communication over commodity internet.}
As in~\cite{gauntlet}, we utilize object storage (specifically Cloudflare R2) as
the communication backbone. This has two benefits: (1) it facilitates validation
of the participants' pseudo-gradients without needing to directly write to the
blockchain; (2) it is particularly synergistic with SparseLoCo, which requires an
all-gather operation over the small pseudo-gradients because uploads are fast
and Cloudflare rapidly distributes the pseudo-gradients, facilitating download
by all other peers.
Participants are required to upload their pseudo-gradients to their Cloudflare
R2 object storage (with the location visible to all participants on the network)
and provide credentials to the storage bucket. The validator then asynchronously selects the
top-scoring pseudo-gradients, and all participants download them directly
from these buckets for aggregation and the global model step. This design avoids
requiring direct peer-to-peer connectivity and supports internet-scale
variability: peers can upload asynchronously, and the validator can fetch,
verify, and score submissions without a synchronized collective.

\paragraph{Bittensor blockchain.}
\textsc{Covenant-72B}, and specifically Gauntlet (the coordination mechanism),
run on top of the Bittensor blockchain under Subnet 3. This provides the
necessary primitives to coordinate peers and has an extensive community of
participants who contribute compute for various AI-related tasks.

\section{Pre-Training}
\label{sec:training_details}

\subsection{Setup}
\paragraph{Model.}

\textsc{Covenant-72B} is a dense decoder-only Transformer in the LLaMA-3 style~\cite{grattafiori2024llama3herdmodels} with grouped-query attention (GQA)~\cite{ainslie2023gqa}, using 80 layers and width $d_{\text{model}}=8192$.
We use 64 attention heads with 8 key-value (KV) heads, Rotary Position Embedding (RoPE) with base frequency $500{,}000$, a maximum context length of 2048, and tied token embeddings and LM head weights.
Tokenization uses the Gemma 3 SentencePiece tokenizer with vocabulary size $V=262{,}208$~\cite{gemma3}. Model details are summarized in Table~\ref{tab:model}.

\paragraph{Data and preprocessing.}
The training data comprises ${\sim}1.1$T tokens in total, split between the main and annealing phases. The main phase (${\sim}1.09$T tokens) consists of web text from DCLM~\cite{li2024datacomp}, while the annealing phase uses higher-quality data~\cite{zyphra_cookbook, blakeney2024doesdatasparkjoy} (${\sim}14.2$B tokens). Specifically, the annealing phase uses a curated blend of instruction (${\sim}27\%$), synthetic web (${\sim}20\%$), code (${\sim}15\%$), math (${\sim}13\%$), and ${\sim}25\%$ pre-training replay data from natural web text to mitigate forgetting (the effect of annealing on downstream benchmarks is summarized in Appendix~\ref{app:anneal}, Table~\ref{tab:anneal-bench}).
To minimize data loading overhead and ensure consistent shard assignment for validation, we pre-tokenize all data and host shards on object storage. Peers download shards ahead of time, replacing consumed shards in the background to avoid on-the-fly tokenization bottlenecks.

\paragraph{Optimization Hyperparameters \& Pseudo-gradient Compression.}
Peers use SparseLoCo~\cite{sparseloco} with AdamW as the inner optimizer and $H=30$ inner steps per training round, with a per-peer batch size of $192$ and sequence length of $2048$. SparseLoCo uses error-feedback decay $\beta{=}0.95$, a constant learning rate of $\alpha{=}1$ for the outer optimizer and pseudo-gradient compression with chunk-wise Top-$k$ sparsification using a chunk size $C{=}4096$ and Top-$k{=}64$, and 2-bit quantization of transmitted values. This leads to a compression ratio of more than $146\times$ relative to dense gradient communication.
The inner AdamW optimizer primarily uses a cosine decay schedule (further described below) with a peak learning rate of $1.2\times 10^{-4}$, weight decay $0.1$, and betas $(0.9,0.95)$.

\begin{figure}[h]
    \centering
    \includegraphics[width=\linewidth]{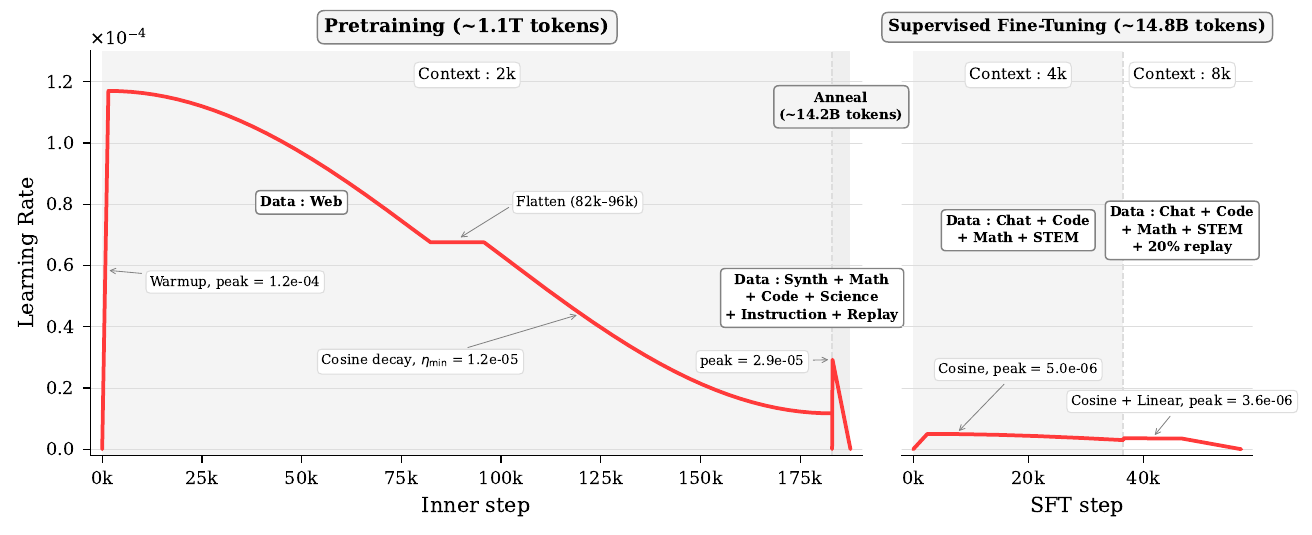}
    \caption{\textbf{Learning rate schedule.} \emph{Left:} pre-training inner
        learning rate with linear warmup, cosine decay with a flatten window,
        followed by an annealing phase on higher-quality data. The cosine decay
        was flattened due to lower participation, which required a longer decay horizon.
        \emph{Right:} Supervised fine-tuning schedule with a 4k-context cosine stage
        followed by an 8k-context cosine-then-linear stage. }
    \label{fig:lr_schedule}
\end{figure}

\paragraph{Inner learning rate schedule.} Figure~\ref{fig:lr_schedule} shows the inner learning rate over the course of training. AdamW's cosine decay schedule uses a peak learning rate of $1.2\times 10^{-4}$, with a linear warmup of $1{,}500$ inner steps (corresponding to $50$ outer steps), decaying the learning rate to a final value of $1.2\times 10^{-5}$. The inner learning rate was flattened for a period towards the middle of training (around the 80K inner-step mark) because, based on the observed number of peers, a longer schedule was needed to reach the target token budget. Thus the inner learning rate was kept flat for $13{,}500$ steps and resumed the decay normally.
During the late stages of training (110K inner steps), we observed that the loss and a number of metrics began to plateau. We found that decreasing the outer learning rate to $\alpha=0.65$ helped alleviate this. In the annealing phase, the inner learning rate is warmed up and rapidly decayed on a higher-quality data mixture to prepare the model for the SFT phase, following~\cite{zyphra_cookbook}.
Finally, we perform a 14.8B-token SFT phase offline, which is further described in Section~\ref{sec:sft}.

\subsection{Main Pre-Training Results}
\label{sec:benchmarks}

\begin{table}[t]
    \centering
    \small
    \setlength{\tabcolsep}{6pt}
    \renewcommand{\arraystretch}{1.18}
    \caption{\textbf{Pre-training benchmark comparison across centrally trained
            and decentralized baselines.} We report zero-shot accuracy on ARC-C,
        ARC-E, PIQA, OpenBookQA, HellaSwag, WinoGrande, and MMLU. Among
        decentralized methods communicating over-the-internet,
        \textsc{Covenant-72B} uses permissionless participation, while INTELLECT-1
        and Psyche Consilience use whitelisted participants; LLM360 K2 and the
        LLaMA models are trained in centralized clusters.
    }
    \label{tab:pretrain-bench}

    \resizebox{\textwidth}{!}{%
        \begin{tabular}{l c c G c c c}
            \toprule
                                   & INTELLECT-1 & Psyche Consilience & Covenant-72B & LLM360 K2   & LLaMA-2-7B  & LLaMA-2-70B \\
            \midrule
            Model size             & 10B         & 40B                & 72B          & 65B         & 7B          & 70B         \\
            Tokens                 & 1T          & 1.2T               & 1.1T         & 1.4T        & 2T          & 2T          \\
            Training env.          & Internet    & Internet           & Internet     & Centralized & Centralized & Centralized \\
            Permissionless         & No          & No                 & Yes          & No          & No          & No          \\
            \addlinespace[2pt]
            \midrule
            \rowcolor{black!5}
            \multicolumn{7}{c}{\textbf{Benchmarks (0-shot accuracy)}}                                                          \\
            \midrule
            \textbf{ARC-Challenge} & 44.8        & 31.1               & 56.8         & 53.8        & 45.1        & 57.4        \\
            \textbf{ARC-Easy}      & 71.8        & 55.8               & 80.9         & 76.0        & 73.8        & 79.6        \\
            \textbf{PIQA}          & 77.4        & 76.1               & 81.6         & 82.5        & 78.7        & 82.6        \\
            \textbf{OpenBookQA}    & 43.8        & 35.2               & 44.0         & 48.0        & 44.2        & 49.4        \\
            \textbf{HellaSwag}     & 70.3        & 63.7               & 80.6         & 82.9        & 76.2        & 84.3        \\
            \textbf{WinoGrande}    & 63.3        & 57.0               & 75.9         & 76.4        & 69.4        & 80.4        \\
            \textbf{MMLU}          & 32.7        & 24.2               & 67.1         & 65.5        & 41.7        & 65.6        \\
            \bottomrule
        \end{tabular}%
    }
\end{table}

We report the final zero-shot benchmark results after pre-training in
Table~\ref{tab:pretrain-bench}, using ARC-Challenge/Easy~\cite{arc},
PIQA~\cite{piqa}, OpenBookQA~\cite{OpenBookQA2018}, HellaSwag~\cite{hellaswag},
WinoGrande~\cite{winogrande}, and MMLU~\cite{mmlu}. We compare to two existing
whitelisted decentralized training efforts at smaller scale as well as two
open-source models of similar size (LLM360 K2 Diamond and LLaMA-2-70B). We also
include LLaMA-2-7B as a reference point. To our knowledge, many
existing open efforts for globally distributed LLM training besides INTELLECT-1
are unable to achieve strong performance and compute utilization, while satisfying the bandwidth
constraints of globally distributed training. We briefly summarize the baseline
models and their training details below. For consistency, we evaluate publicly
available checkpoints across benchmarks using~\cite{eval-harness} under a
unified evaluation protocol (details in Appendix~\ref{app:eval-setup}). All
evaluated checkpoints are hosted on Hugging Face, and we list the exact model
identifiers used below.

\paragraph{INTELLECT-1.}
INTELLECT-1~\cite{intellect1} is a permissioned globally distributed pre-training run that trained a
10B-parameter dense Transformer LLM over 1T tokens.
Training used PRIME, combining DiLoCo with int8 all-reduce to reduce
cross-node communication, while supporting dynamic node participation (up to 14 nodes / 112 H100s). We evaluate the Hugging Face checkpoint \texttt{PrimeIntellect/INTELLECT-1} for Table~\ref{tab:pretrain-bench}.

\paragraph{Psyche Consilience.}
Psyche Consilience~\cite{psyche-consilience-40b-hf} is another ongoing whitelisted decentralized pre-training run that trains a
40B-parameter dense decoder-only LLM.
Consilience uses a communication-efficient single-step optimizer, DeMo~\cite{demo}, and is trained on a mixture of FineWeb, FineWeb-2, and The Stack v2. We evaluate the checkpoint from the first run \texttt{PsycheFoundation/consilience-40b-7Y9v38s5}~\cite{psyche-consilience-40b-hf} for Table~\ref{tab:pretrain-bench}.

\paragraph{LLM360 K2.}
LLM360 K2 Diamond~\cite{liu2025llm360} is a 65B-parameter dense Transformer pre-trained in a conventional
centralized-cluster setting using AdamW. Relative to our setting, K2 provides a strong
centralized baseline near the same parameter scale, and a slightly larger token budget. In Table~\ref{tab:pretrain-bench}, we evaluate this model using the Checkpoint 360 from the Hugging Face repository~\texttt{LLM360/K2}.

\paragraph{LLaMA-2.}
LLaMA-2-70B~\cite{touvron2023llama} is a 70B-parameter dense decoder-only Transformer pre-trained by Meta in a conventional centralized-cluster setting. It is trained on 2T tokens with a 4k context window; the 70B variant uses grouped-query attention (GQA) while the 7B variant does not.
We include LLaMA-2-70B as a strong datacenter-trained baseline at a similar parameter count and
architecture, but trained on nearly $2\times$ as many tokens (2T vs.\ ${\sim}1.1$T).
In Table~\ref{tab:pretrain-bench} we evaluate LLaMA-2 models using the publicly available checkpoints \texttt{meta-llama/Llama-2-7b-hf} and \texttt{meta-llama/Llama-2-70b-hf}.

\textsc{Covenant-72B} is substantially larger in scale (in terms of size and compute) than existing training runs over globally distributed compute and far exceeds the performance of prior decentralized models. Across all reported tasks, \textsc{Covenant-72B} achieves competitive downstream
performance compared to centralized baselines despite being trained over commodity internet links with
permissionless participation, demonstrating that large-scale collaborative
pre-training can reach competitive quality without relying on whitelisting or
centralized datacenter training environments. Specifically, we observe stronger
performance in ARC-Challenge, MMLU, and ARC-Easy than K2, and exceeding or on par with LLaMA-2-70B.
Improvements in these metrics were also observed in small-scale experiments
compared with AdamW training on the same data. We observe slightly lower
performance across HellaSwag, OpenBookQA, and WinoGrande than K2 and
LLaMA-2-70B, which were trained on larger token budgets. We hypothesize that these differences are primarily driven by
dissimilarities in data quality/mixture and training recipes rather than
infrastructure, and suggest that SparseLoCo and other low-bandwidth optimization methods are able to scale to the largest-scale training tasks. Finally, we observe that Covenant-72B well exceeds the performance of smaller-scale and other
decentralized models.

Overall, \emph{compared to centralized-cluster training runs of similar parameter count, \textsc{Covenant-72B} is broadly competitive.} Notably, these centralized baselines were trained with conventional datacenter infrastructure and, in the case of LLaMA-2-70B, on substantially more tokens (2T vs.\ ${\sim}1.1$T). Although these comparisons are not fully controlled (differences in data mixtures, tokenizers, training recipes, and token budgets), Table~\ref{tab:pretrain-bench} suggests that decentralized, permissionless pre-training can approach the quality of standard centralized runs at similar scale.

\subsection{Communication Efficiency}
\label{sec:timing}

In local optimizers such as DiLoCo/SparseLoCo~\cite{diloco,sparseloco,streamingdiloco,therien2025muloco,obeidi2026heterogeneous}, each training round consists of (i) a \emph{compute phase}, where each peer runs $H$ inner-optimizer steps from the same global model, and (ii) a \emph{communication phase}, covering everything else such as pseudo-gradient preparation, compression, synchronization, aggregation, and the outer optimizer step that advances all peers' local models to the next shared model. Here, we report the wall-clock time spent in each phase to quantify the communication overhead of collaborative internet-scale training.

With $R{=}20$ peers, $H{=}30$ inner steps per round, and $8\times$B200 per peer, we enforce a fixed per-round compute window of $t_{\text{compute}}{=}20$ minutes. Assuming a bandwidth constraint where each node does not exceed 500\,Mb/s downlink and 110\,Mb/s uplink, we observe an average communication time of $t_{\text{comm}}{=}70$ seconds per round. This corresponds to a compute utilization of ${\sim}94.5\%$ for the 72B model.

\begin{figure}[h!]
    \centering
    \includegraphics[width=\linewidth]{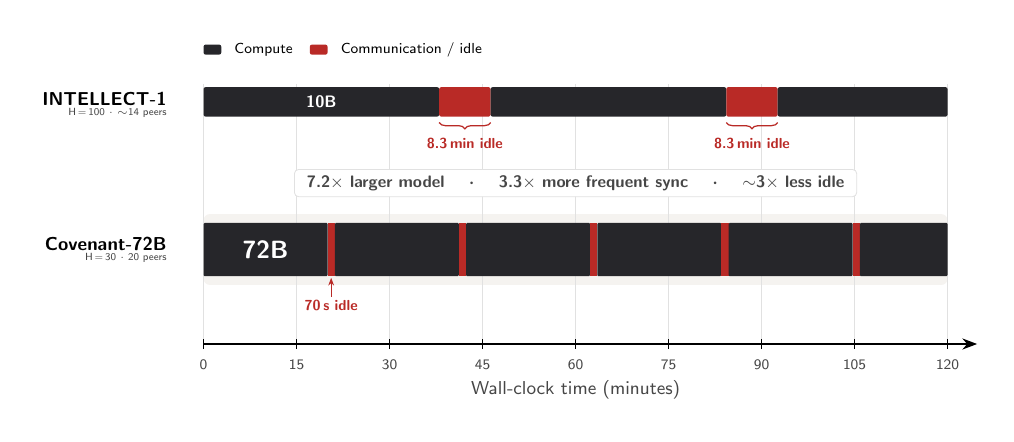}
    \caption{\textbf{Compute--communication timelines over a two-hour window.}
        Each row shows the breakdown of successive training rounds, with black segments denoting the compute window (inner-step training) and red segments denoting synchronization overhead. Despite training a $7.2\times$ larger model, \textsc{Covenant-72B} incurs only 70\,s of idle time per round, compared to the 8.3\,min per-round synchronization overhead reported for DiLoCo-style training in INTELLECT-1.}

    \label{fig:timeline}
\end{figure}

For context, we compare to the other major globally distributed run INTELLECT-1~\cite{intellect1}, which reports $t_{\text{compute}} \approx 38$ minutes for $H{=}100$ inner steps, $8\times$H100 per peer when training a 10B model. Moreover, they report $t_{\text{comm}} \approx 8.3$ minutes on average for synchronization at a peak configuration of ${\sim}14$ nodes. This corresponds to an ${\sim}82.1\%$ compute utilization. Notably, synchronization is performed every $H{=}100$ steps in this setting ($\approx 3.33\times$ less frequently) which comes with performance degradation. In a more direct comparison, SparseLoCo~\cite{sparseloco} reports for an 8B model with $R{=}15$ peers, $H{=}30$ inner steps, and $8\times$H200 per peer, an average communication time of $t_{\text{comm}} \approx 12$ seconds under 500\,Mb/s downlink and 110\,Mb/s uplink bandwidth constraints. With a computation time of $t_{\text{compute}} \approx 4.5$ minutes, this yields a compute utilization of ${\sim}95.7\%$. Figure~\ref{fig:timeline} visualizes the training round structure over a two-hour window, highlighting the difference in idle time between the two systems.

\subsection{Participation Dynamics}
\label{sec:participation}

\begin{figure}[t]
    \centering
    \begin{minipage}[t]{0.54\linewidth}
        \centering
        \includegraphics[width=\linewidth]{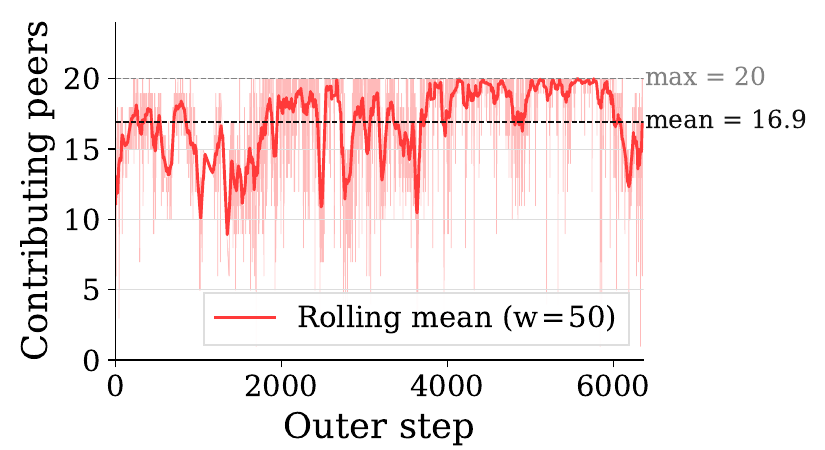}
        \caption{\textbf{Contributing peers over the course of training.}
            The solid curve shows the number of peers whose pseudo-gradients were selected (by Gauntlet) and included in each round's aggregation. We cap the number of contributors at 20; across the run, we observed an average of 16.9 contributing peers throughout training.}

        \label{fig:contributing_peers}
    \end{minipage}
    \hfill
    \begin{minipage}[t]{0.42\linewidth}
        \centering
        \includegraphics[width=\linewidth]{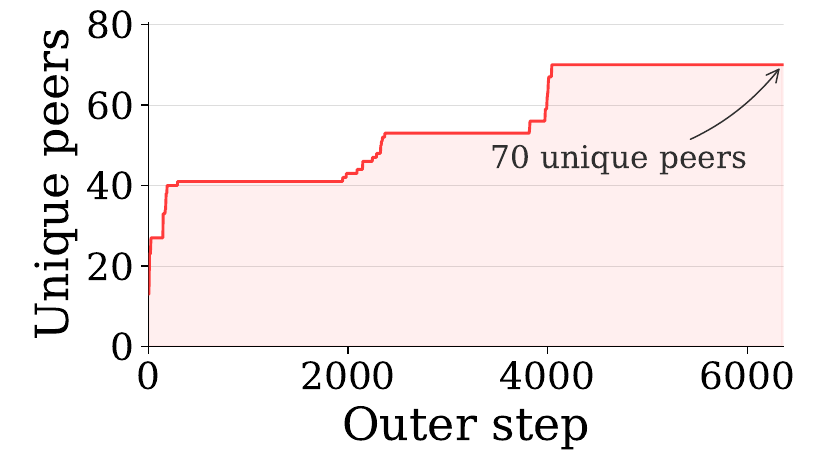}
        \caption{\textbf{Cumulative unique peer participants over training.} At
            least 70 unique peers contributed to model updates over the course of
            the run. }
        \label{fig:unique_peers}
    \end{minipage}
\end{figure}

In decentralized training, peer participation can be dynamic as participants join and leave at their discretion or due to unexpected circumstances. Figure~\ref{fig:contributing_peers} shows the number of contributing peers per round over the entire run. Despite this dynamism, participation remains close to the maximum of 20 throughout, with a mean of 16.9 contributing peers, and SparseLoCo is robust to this fluctuation. This is due in part to the calibration of the reward mechanism that incentivizes new participants to join quickly once others leave.

Figure~\ref{fig:unique_peers} shows the cumulative number of unique peer IDs observed during training through analysis of the blockchain. Because UIDs registered on the Bittensor blockchain can change ownership over time, and we track only UIDs, the reported count is a lower bound on the true number of distinct participants. We report further details on the number of active and contributing peers in Appendix~\ref{app:participation}.

\section{Covenant-72B-Chat: Supervised Fine-Tuning (SFT)}
\label{sec:sft}

After pre-training, we fine-tune on ${\sim}14.8$B tokens in two stages to produce \textsc{Covenant-72B-Chat},
progressively extending the effective context length from the 2048-token pre-training window and making the model suitable for interaction.

\paragraph{Data.}
Our instruction dataset draws from open conversation and instruction-following
collections~\cite{smoltalk} as well as post-training data spanning chat, code, math, STEM,
competitive programming, and agentic tasks. We keep only non-reasoning examples with at
least two messages per conversation, and format everything with a chat template using
\texttt{<start\_of\_turn>}/\texttt{<end\_of\_turn>} delimiters and the same tokenizer used in
pre-training. We prepare two variants of the dataset, truncated to 4096 and 8192 tokens, respectively. For the 8k variant, we additionally mix in 20\% pre-training replay data sampled from natural web
text, shuffled uniformly into the instruction data. This helps prevent regression on pre-trained capabilities during fine-tuning.

\textbf{Stage~1: 4k context.}
Starting from the pre-trained \textsc{Covenant-72B} checkpoint, we fine-tune on the 4k data for
36{,}500 steps (${\sim}68\%$ of one epoch) with a global batch size of 256 and a maximum sequence
length of 4096. Sequences are variable-length (no packing), handled via nested tensors. We use AdamW
with a peak learning rate of $5 \times 10^{-6}$, betas $(0.9,\, 0.95)$, weight decay $0.01$, and
gradient clipping at $1.0$, under a cosine schedule spanning 1.5 epochs with 3\% warmup. Training runs in bfloat16 with
FSDP2, gradient checkpointing, and \texttt{torch.compile}.

\textbf{Stage~2: 8k context with replay.}
We continue from the first stage's checkpoint on the 8k data (which includes the 20\% pre-training replay),
extending the maximum sequence length to 8192. To keep the transition smooth, we
initialize the learning rate where the previous stage's cosine schedule left off (${\approx}\,2.97 \times 10^{-6}$),
warm up over 25 steps to a peak of $3.57 \times 10^{-6}$, and follow a cosine schedule until step 10{,}100
before switching to linear decay to zero over the remaining 10{,}400 steps (20{,}500 total).
All other optimizer settings carry over from Stage~1.
These learning rate schedules are further illustrated in Figure~\ref{fig:lr_schedule}.

\paragraph{Quantitative Results.} Table~\ref{tab:chat-5shot-bench} shows the results of
standard 5-shot evaluations on the post-SFT models, using the benchmarks
in pre-training (Table~\ref{tab:pretrain-bench}) as well as additional benchmarks including GSM8K~\cite{cobbe2021gsm8k}, BBH
CoT~\cite{suzgun2022bbh},
IFEval~\cite{zhou2023ifeval}, MATH~\cite{hendrycks2021math},
MMLU-Pro~\cite{wang2024mmlupro}, and MuSR~\cite{sprague2024musr}. These additional benchmarks are typically challenging for base models, and we see significant progress in them from the SFT. To align with literature, we use 25-shot for ARC-Challenge and 10-shot for HellaSwag. For BBH, we use 3-shot as in~\cite{suzgun2022bbh}, and for MATH, we use 4-shot as in~\cite{lewkowycz2022minerva}. We primarily
compare \textsc{Covenant-72B-Chat} with centralized-cluster trained models
K2-Chat~\cite{liu2025llm360} and LLaMA-2-70B-Chat~\cite{touvron2023llama} using
their SFT checkpoints \texttt{LLM360/K2-Chat} and
\texttt{meta-llama/Llama-2-70b-chat-hf}, respectively. Compared to K2-Chat and
LLaMA-2-70B-Chat, we observe competitive metrics in most categories.
Notably,
\textsc{Covenant-72B-Chat} achieves the highest IFEval and MATH scores among all
compared models, suggesting strong instruction-following and mathematical
reasoning capabilities after SFT. The chat model also retains strong performance
on the same benchmarks used for pre-training evaluation
(Table~\ref{tab:pretrain-bench}), indicating that the two-stage fine-tuning
pipeline, including the 8k context extension and 20\% pre-training replay in
Stage~2, largely preserves or improves the capabilities acquired during pre-training.
Moreover, the model handles a range of standard instruction-following, math,
and coding topics as shown in Appendix~\ref{app:qualitative}.

\begin{table}[t]
    \centering
    \small
    \setlength{\tabcolsep}{6pt}
    \renewcommand{\arraystretch}{1.18}
    \caption{\textbf{Benchmark results on chat models.} Values are accuracy
        (\%) with one decimal. We use 25-shot for ARC-Challenge, 10-shot for HellaSwag, 3-shot for BBH CoT, and 4-shot for MATH; all remaining benchmarks use 5-shot. Metrics are
        \texttt{acc\_norm} where available (except MMLU and WinoGrande
        \texttt{acc}, and GSM8K \texttt{strict}); additional benchmarks use
        \texttt{exact\_match} (BBH CoT, MATH, MMLU-Pro), \texttt{prompt\_strict}
        (IFEval), and \texttt{acc\_norm} (MuSR). }
    \label{tab:chat-5shot-bench}

    \resizebox{\textwidth}{!}{%
        \begin{tabular}{l c c c G}
            \toprule
                                   & LLaMA-2-7B-Chat & LLaMA-2-70B-Chat & K2-Chat (65B) & Covenant-72B-Chat \\
            \midrule
            \rowcolor{black!5}
            \multicolumn{5}{c}{\textbf{Benchmarks}}                                                \\
            \midrule
            \textbf{ARC-Challenge} & 53.2            & 65.4             & 62.0          & 64.2              \\
            \textbf{ARC-Easy}      & 80.6            & 85.3             & 85.8          & 85.5              \\
            \textbf{GSM8K }        & 22.6            & 52.2             & 79.0          & 63.9              \\
            \textbf{HellaSwag}     & 78.6            & 85.9             & 79.3          & 79.2              \\
            \textbf{MMLU}          & 47.2            & 63.1             & 67.9          & 67.4              \\
            \textbf{OpenBookQA}    & 42.6            & 47.4             & 48.2          & 51.8              \\
            \textbf{PIQA}          & 78.2            & 81.6             & 83.4          & 82.8              \\
            \textbf{WinoGrande}    & 72.5            & 79.6             & 79.6          & 77.3              \\
            \midrule
            \rowcolor{black!5}
            \multicolumn{5}{c}{\textbf{Additional Benchmarks}}                                              \\
            \midrule
            \textbf{BBH CoT}       & 40.4            & 63.2             & 69.8          & 55.0              \\
            \textbf{IFEval}        & 30.9            & 40.7             & 45.5          & 64.7              \\
            \textbf{MATH}          & 4.8             & 10.7             & 19.1          & 26.3              \\
            \textbf{MMLU-Pro}      & 22.9            & 35.2             & 45.4          & 40.9              \\
            \textbf{MuSR}          & 40.2            & 48.7             & 46.6          & 39.7              \\
            \bottomrule
        \end{tabular}%
    }
\end{table}

\section{Conclusion}
In this report, we introduced \textsc{Covenant-72B}, a 72B-parameter LLM pre-trained over commodity internet links with \emph{permissionless} participation. By combining the Gauntlet incentivization and validation mechanism with the communication-efficient SparseLoCo optimizer, the run supports peers dynamically joining and leaving while maintaining high utilization and strong end-model quality. Across standard zero-shot evaluations, \textsc{Covenant-72B} is broadly competitive with centralized baselines at similar scale, and substantially improves over prior decentralized runs, suggesting that infrequent pseudo-gradient communication with aggressive compression can enable training at unprecedented scale under real-world networking constraints. We additionally perform supervised fine-tuning (SFT) to obtain \textsc{Covenant-72B-Chat}, which achieves competitive performance compared to similarly sized centrally trained chat models.

Future work can consider scaling training to a wider and potentially more heterogeneous set of participants, as well as exploring alternatives to trustless peer participation. More broadly, \textsc{Covenant-72B} points toward a practical path for \emph{permissionless}, globally distributed training---where open participation, rather than centralized access to tightly coupled infrastructure, becomes the default mechanism for scaling and \emph{democratizing} foundation model training.

\bibliographystyle{plain} 
\bibliography{bibliography} 

\clearpage
\appendix

\section{Participation}\label{app:participation}
\begin{figure}[H]
    \centering
    \includegraphics[width=\linewidth]{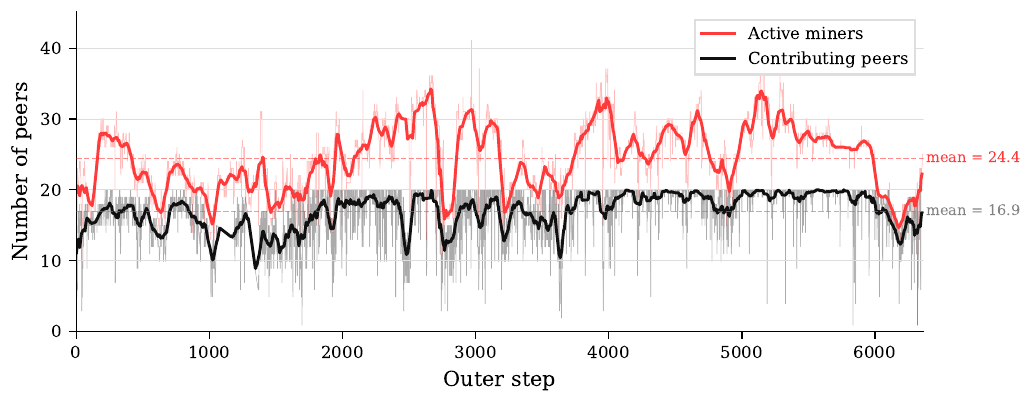}
    \caption{\textbf{Active and contributing peers over training.}
        Active peers (red) are registered on the network and actively submitting pseudo-gradients; contributing peers (black) denote the number of peers whose pseudo-gradients are selected for aggregation each round. In our permissionless setting, not all submissions are selected (e.g., due to failing validation checks or low-quality pseudo-gradients).}
    \label{fig:peer_participation}
\end{figure}

In permissionless decentralized training, peer availability changes over time: participants may join, leave, or pause due to network issues or hardware problems. Figure~\ref{fig:peer_participation} shows the number of peers actively submitting pseudo-gradients per step (red). Because participation is open, we use the Gauntlet mechanism to filter out submissions that appear low-quality or bad-faith (e.g., suspected of copying). The contributing peers (black) are those whose submissions are selected for the final aggregation and model update. Across the run, we observe an average of 24.4 active peers per step and 16.9 contributing peers per step.

\section{Effect of Annealing}\label{app:anneal}

Table~\ref{tab:anneal-bench} compares the base model's performance immediately
before and after the ${\sim}14.2$B-token annealing phase. We can see that some
of the simpler tasks were actually degraded slightly while more complex tasks
were improved. The goal of this phase is also to better prepare the model for
post-training~\cite{zyphra_cookbook}.

\begin{table}[H]
    \centering
    \small
    \setlength{\tabcolsep}{6pt}
    \renewcommand{\arraystretch}{1.18}
    \caption{\textbf{Base model performance before and after annealing.} Zero-shot accuracy on the same benchmarks as Table~\ref{tab:pretrain-bench}. The pre-anneal checkpoint corresponds to step $6{,}100$ (${\sim}1.09$T tokens) and the post-anneal checkpoint to step $6{,}190$ (${\sim}1.1$T tokens).}
    \label{tab:anneal-bench}
    \begin{tabular}{l c c}
        \toprule
                            & Pre-Anneal & Post-Anneal \\
        \midrule
        \textbf{ARC-C}      & 56.4       & 56.8        \\
        \textbf{ARC-E}      & 81.2       & 80.9        \\
        \textbf{PIQA}       & 82.2       & 81.6        \\
        \textbf{OpenBookQA} & 44.4       & 44.0        \\
        \textbf{HellaSwag}  & 81.4       & 80.6        \\
        \textbf{WinoGrande} & 75.7       & 75.9        \\
        \textbf{MMLU}       & 62.5       & 67.1        \\
        \bottomrule
    \end{tabular}
\end{table}

\section{Model Details}\label{app:model}
Table~\ref{tab:model} lists the model and tokenizer configuration.
\begin{table}[H]
    \centering
    \setlength{\tabcolsep}{6pt}
    \caption{\textbf{Model configuration for \textsc{Covenant-72B}.}}
    \label{tab:model}
    \begin{tabular}{@{}ll@{}}
        \toprule
        \textbf{Field}  & \textbf{Value}       \\
        \midrule
        Parameters      & 72{,}747{,}327{,}488 \\
        Layers          & 80                   \\
        Model width     & 8192                 \\
        Query heads     & 64                   \\
        KV heads        & 8                    \\
        RoPE ($\theta$) & $500{,}000$          \\
        Tokenizer       & Gemma 3              \\
        Vocab Size      & $262{,}208$          \\
        \bottomrule
    \end{tabular}
\end{table}

\section{Evaluation Setup}\label{app:eval-setup}

All benchmarks reported in this paper are evaluated using \texttt{lm-eval}
v0.4.11~\cite{eval-harness} with the \texttt{vllm} v0.16.0 inference backend,
running \texttt{torch} 2.9.1 and \texttt{transformers} 4.57.6. The one exception
is Psyche Consilience, whose dense DeepSeek-v3 architecture is incompatible with
\texttt{vllm}. For this model we use the Hugging Face backend with
\texttt{accelerate} v1.13.0.

For the SFT evaluations (Table~\ref{tab:chat-5shot-bench}), we additionally
use \texttt{math-verify} v0.9.0 for MATH scoring. The LLaMA-2 chat models
are evaluated without \texttt{-{}-apply\_chat\_template} because LLaMA-2's chat
template enforces strict alternating user/assistant roles, which is incompatible
with standard few-shot prompt construction. K2-Chat is evaluated with
\texttt{-{}-apply\_chat\_template default}, using the named template from its
\texttt{tokenizer\_config.json}. K2 and INTELLECT-1 base checkpoints require
\texttt{add\_bos\_token=True} to match their training configuration.

\section{Qualitative Examples from Covenant-72B-Chat}\label{app:qualitative}

Below we present selected prompts and corresponding responses generated by \textsc{Covenant-72B-Chat} across several task categories. These examples are included to give a qualitative sense of the model's capabilities and failure modes after supervised fine-tuning. All responses are reproduced verbatim (including errors).

\clearpage
\subsection{Math Reasoning}

\begin{tcolorbox}[colback=covenantwhite, colframe=covenantred, fonttitle=\bfseries, title=Prompt]
    \small Without listing them all, find the sum of the integers from 1 to 100 that are divisible by 3 or 5.
\end{tcolorbox}

\begin{tcolorbox}[colback=white, colframe=covenantblk, fonttitle=\bfseries, title=Response]
    \small\setlength{\parskip}{0.5em}\setlength{\parindent}{0pt}
    To find the sum of the integers from 1 to 100 that are divisible by 3 or 5, we can use the principle of inclusion-exclusion. This principle states that when we want to find the sum of elements in the union of two sets, we add the sums of the individual sets and then subtract the sum of their intersection to avoid double counting.

    First, let's find the sum of the integers from 1 to 100 that are divisible by 3. The largest multiple of 3 less than or equal to 100 is 99, so we have 33 multiples of 3. The sum of an arithmetic series is given by the formula n(a + l)/2, where n is the number of terms, a is the first term, and l is the last term. In this case, n = 33, a = 3, and l = 99. Therefore, the sum of the multiples of 3 is 33(3 + 99)/2 = 33(102)/2 = 1683.

    Next, let's find the sum of the integers from 1 to 100 that are divisible by 5. The largest multiple of 5 less than or equal to 100 is 100, so we have 20 multiples of 5. Using the same formula, the sum of the multiples of 5 is 20(5 + 100)/2 = 20(105)/2 = 1050.

    Now, we need to find the sum of the integers from 1 to 100 that are divisible by both 3 and 5, i.e., the multiples of 15. The largest multiple of 15 less than or equal to 100 is 90, so we have 6 multiples of 15. Using the formula again, the sum of the multiples of 15 is 6(15 + 90)/2 = 6(105)/2 = 315.

    Finally, we apply the principle of inclusion-exclusion: the sum of the integers from 1 to 100 that are divisible by 3 or 5 is the sum of the multiples of 3 plus the sum of the multiples of 5 minus the sum of the multiples of 15. Therefore, the sum is 1683 + 1050 - 315 = 2418.
\end{tcolorbox}

\begin{tcolorbox}[colback=covenantwhite, colframe=covenantred, fonttitle=\bfseries, title=Prompt]
    \small Consider the identity: $9 \times 4 + 6 \times 6 = 72$. Can you modify exactly one integer (and not more than that!) on the left hand side of the equation so the right hand side becomes 99? Let's think step by step, write down a plan, and then write down your solution as: ``The solution is: $A \times B + C \times D$''.
\end{tcolorbox}

\begin{tcolorbox}[colback=white, colframe=covenantblk, fonttitle=\bfseries, title=Response]
    \small\setlength{\parskip}{0.5em}\setlength{\parindent}{0pt}
    To solve this problem, we need to modify exactly one integer on the left-hand side of the equation to make the right-hand side equal to 99.

    Here's a plan to approach this problem:

    We start by analyzing the given equation: $9 \times 4 + 6 \times 6 = 72$. We can see that the left-hand side consists of two multiplication operations followed by an addition operation. To make the right-hand side equal to 99, we need to modify exactly one integer in the left-hand side.

    Since the goal is to make the right-hand side equal to 99, we can try modifying the first multiplication operation, $9 \times 4$, to see if it results in the desired outcome.

    We can try replacing 4 with a different integer, say $x$. The modified equation would then be: $9 \times x + 6 \times 6$. We want the right-hand side to be equal to 99.

    Since $6 \times 6$ equals 36, we can set up an equation: $9 \times x + 36 = 99$.

    Now, let's solve for $x$:

    $9 \times x = 99 - 36$\\
    $9 \times x = 63$\\
    $x = 63 / 9$\\
    $x = 7$

    So, replacing 4 with 7 in the first multiplication operation results in the equation $9 \times 7 + 6 \times 6$, which equals 99.

    The solution is: $9 \times 7 + 6 \times 6$.
\end{tcolorbox}

\subsection{Logical Reasoning}

\begin{tcolorbox}[colback=covenantwhite, colframe=covenantred, fonttitle=\bfseries, title=Prompt]
    \small Three friends---Alice, Ben, and Cara---each own a different pet: a cat, dog, or turtle. Alice does not own the cat. Ben does not own the dog. Cara owns neither the dog nor the turtle. Who owns which pet?
\end{tcolorbox}

\begin{tcolorbox}[colback=white, colframe=covenantblk, fonttitle=\bfseries, title=Response]
    \small\setlength{\parskip}{0.5em}\setlength{\parindent}{0pt}
    Let's break down the problem step by step:

    We have three friends: Alice, Ben, and Cara. Each owns a different pet: a cat, a dog, or a turtle. The constraints are:

    \begin{enumerate}[nosep,leftmargin=*]
        \item Alice does not own the cat.
        \item Ben does not own the dog.
        \item Cara owns neither the dog nor the turtle.
    \end{enumerate}

    \textbf{Step 1: Analyze Cara's Pet}

    Cara owns neither the dog nor the turtle, so Cara must own the \textbf{cat}. This is because the only remaining pet is the cat, and Cara cannot own the dog or turtle.

    \textbf{Step 2: Determine Remaining Pets}

    Since Cara owns the cat, the remaining pets to be owned by Alice and Ben are the \textbf{dog} and the \textbf{turtle}.

    \textbf{Step 3: Apply Constraints to Alice and Ben}

    \begin{itemize}[nosep,leftmargin=*]
        \item \textbf{Alice does not own the cat} (already satisfied since Cara owns the cat).
        \item \textbf{Ben does not own the dog} (so Ben cannot own the dog).
    \end{itemize}

    This leaves:

    \begin{itemize}[nosep,leftmargin=*]
        \item \textbf{Alice} must own the \textbf{dog} (since Ben cannot own the dog).
        \item \textbf{Ben} must own the \textbf{turtle} (since the only remaining pet is the turtle).
    \end{itemize}

    \textbf{Final Assignment:}

    \begin{itemize}[nosep,leftmargin=*]
        \item \textbf{Alice}: Dog
        \item \textbf{Ben}: Turtle
        \item \textbf{Cara}: Cat
    \end{itemize}

    \textbf{Verification:}

    \begin{itemize}[nosep,leftmargin=*]
        \item Alice does not own the cat $\checkmark$
        \item Ben does not own the dog $\checkmark$
        \item Cara owns neither the dog nor the turtle $\checkmark$ (she owns the cat)
    \end{itemize}

    \textbf{Final Answer:}

    Alice owns the dog, Ben owns the turtle, and Cara owns the cat.

    \textbf{Answer:}

    Alice owns the dog, Ben owns the turtle, and Cara owns the cat.
\end{tcolorbox}

\begin{tcolorbox}[colback=covenantwhite, colframe=covenantred, fonttitle=\bfseries, title=Prompt]
    \small You have three light switches outside a closed room, and each controls one of three bulbs inside. You may enter the room only once. How do you figure out which switch controls which bulb?
\end{tcolorbox}

\begin{tcolorbox}[colback=white, colframe=covenantblk, fonttitle=\bfseries, title=Response]
    \small\setlength{\parskip}{0.5em}\setlength{\parindent}{0pt}
    To solve this problem, we need to devise a strategy that allows us to gather the most information possible from a single observation. Here's a step-by-step approach:

    First, turn switch 1 to the "on" position for 5 minutes. Then, turn it off. Next, turn switch 2 to the "on" position and immediately enter the room.

    Now, observe the bulbs. One of them will be on, one will be off but warm, and one will be off and cold. The bulb that is on is controlled by switch 2, since we turned that switch on just before entering the room.

    The bulb that is off but warm is controlled by switch 1, which we had turned on for 5 minutes before turning it off. This allowed the corresponding bulb to heat up.

    The remaining bulb, which is off and cold, must be controlled by switch 3, which we never turned on.

    This approach allows us to figure out which switch controls which bulb with just one observation, by using the heat of the bulbs as an additional piece of information.
\end{tcolorbox}

\begin{tcolorbox}[colback=covenantwhite, colframe=covenantred, fonttitle=\bfseries, title=Prompt]
    \small Dana, Eli, and Farah finished a race. Dana finished ahead of Eli. Farah did not finish last. Eli was not first. Dana did not finish first. What was the finishing order?
\end{tcolorbox}

\begin{tcolorbox}[colback=white, colframe=covenantblk, fonttitle=\bfseries, title=Response]
    \small\setlength{\parskip}{0.5em}\setlength{\parindent}{0pt}
    To determine the correct finishing order, let's analyze the given information step by step:
    \begin{enumerate}[nosep,leftmargin=*]
    \item Dana finished ahead of Eli.
    \item Farah did not finish last.
    \item Eli was not first.
    \item Dana did not finish first.
    \end{enumerate}

    From point 1, we know that Dana finished ahead of Eli, which means Dana's position is higher than Eli's.

    From point 2, we know that Farah did not finish last, which means Farah's position is not the lowest.

    From point 3, we know that Eli was not first, which means Eli's position is not the highest.

    From point 4, we know that Dana did not finish first, which means Dana's position is not the highest.

    Now, let's combine the information:

    Since Dana did not finish first (point 4) and Eli was not first (point 3), the highest position must be occupied by Farah.

    Now we have:

    1. Farah
    2. Dana
    3. Eli

    We still need to determine the correct order between Dana and Eli.

    From point 1, we know that Dana finished ahead of Eli, which means Dana's position is higher than Eli's.

    So, the correct finishing order is:
    
    1. Farah
    2. Dana
    3. Eli
\end{tcolorbox}

\subsection{Planning}

\begin{tcolorbox}[colback=covenantwhite, colframe=covenantred, fonttitle=\bfseries, title=Prompt]
    \small Plan a one-day workshop for 25 researchers on efficient LLM training. Include sessions, breaks, and a closing discussion.
\end{tcolorbox}

\begin{tcolorbox}[colback=white, colframe=covenantblk, fonttitle=\bfseries, title=Response]
    \small\setlength{\parskip}{0.5em}\setlength{\parindent}{0pt}
     Here's a structured plan for a one-day workshop on efficient LLM training for 25 researchers: \vspace{-1mm}

    \textbf{Workshop Agenda: Efficient LLM Training\vspace{-1mm}}

    \textbf{9:00 AM – 9:30 AM: Registration \& Welcome\vspace{-1mm}}
    \begin{itemize}[nosep,leftmargin=*]
        \item \textbf{Activity:} Participants check in, grab coffee/snacks, and mingle.
        \item \textbf{Facilitator:} Welcome speech and introduction to the day’s goals.
        \item \textbf{Key Message:} Emphasize the importance of efficiency in LLM training for sustainability and scalability. \vspace{-1mm}
    \end{itemize}

    \textbf{9:30 AM – 10:30 AM: Key Concepts in Efficient LLM Training\vspace{-1mm}}
    \begin{itemize}[nosep,leftmargin=*]
        \item \textbf{Session:}
              \begin{itemize}[nosep,leftmargin=*]
                  \item Overview of efficiency metrics (e.g., FLOPs, energy, training time).
                  \item Case studies: Efficient models (e.g., LLaMA, Mistral, TinyLLM).
                  \item Challenges in scaling LLMs (e.g., data, hardware, computational resources).
              \end{itemize}
        \item \textbf{Facilitator:} Interactive Q\&A to clarify concepts.
        \item \textbf{Takeaway:} Participants understand the trade-offs between model size, accuracy, and efficiency. \vspace{-1mm}
    \end{itemize}

    \textbf{10:30 AM – 11:00 AM: Break\vspace{-1mm}}
    \begin{itemize}[nosep,leftmargin=*]
        \item \textbf{Activity:} Networking, coffee, and light snacks. \vspace{-1mm}
    \end{itemize}

    \textbf{11:00 AM – 12:00 PM: Practical Strategies for Efficient Training\vspace{-1mm}}
    \begin{itemize}[nosep,leftmargin=*]
        \item \textbf{Session:}
              \begin{itemize}[nosep,leftmargin=*]
                  \item \textbf{Model Compression:} Quantization, pruning, and knowledge distillation.
                  \item \textbf{Hardware Optimization:} GPU/TPU utilization, distributed training, and cloud vs. on-premise solutions.
                  \item \textbf{Data Efficiency:} Data augmentation, transfer learning, and active learning.
              \end{itemize}
        \item \textbf{Facilitator:} Live demo of a compressed model or distributed training setup.
        \item \textbf{Takeaway:} Actionable techniques for reducing resource usage without sacrificing performance. \vspace{-1mm}
    \end{itemize}

    \textbf{12:00 PM – 1:00 PM: Lunch Break\vspace{-1mm}}
    \begin{itemize}[nosep,leftmargin=*]
        \item \textbf{Activity:} Network over lunch. \vspace{-1mm}
    \end{itemize}

    \textbf{1:00 PM – 2:00 PM: Case Studies \& Best Practices\vspace{-1mm}}
    \begin{itemize}[nosep,leftmargin=*]
        \item \textbf{Session:}
              \begin{itemize}[nosep,leftmargin=*]
                  \item \textbf{Case Study 1:} Efficient training for a small-scale LLM (e.g., 1B parameters).
                  \item \textbf{Case Study 2:} Scaling an LLM to 10B parameters while maintaining efficiency.
                  \item \textbf{Discussion:} Group activity to brainstorm challenges and solutions.
              \end{itemize}
        \item \textbf{Facilitator:} Facilitate debate and share insights from industry experts.
        \item \textbf{Takeaway:} Participants gain real-world examples and learn from others’ experiences. \vspace{-1mm}
    \end{itemize}

    \textbf{2:00 PM – 2:15 PM: Break\vspace{-1mm}}
    \begin{itemize}[nosep,leftmargin=*]
        \item \textbf{Activity:} Refreshments and informal networking. \vspace{-1mm}
    \end{itemize}

    \textbf{2:15 PM – 3:15 PM: Advanced Topics in Efficiency\vspace{-1mm}}
    \begin{itemize}[nosep,leftmargin=*]
        \item \textbf{Session:}
              \begin{itemize}[nosep,leftmargin=*]
                  \item \textbf{Emerging Trends:} Self-supervised learning, few-shot learning, and energy-aware training.
                  \item \textbf{Ethical Considerations:} Environmental impact, bias in efficient models, and transparency.
                  \item \textbf{Tools \& Frameworks:} Open-source tools for efficiency tracking (e.g., TensorBoard, LLM-specific benchmarks).
              \end{itemize}
        \item \textbf{Facilitator:} Panel-style discussion with experts on cutting-edge research.
        \item \textbf{Takeaway:} Awareness of future trends and ethical responsibilities. \vspace{-1mm}
    \end{itemize}

    \textbf{3:15 PM – 3:30 PM: Break\vspace{-1mm}}
    \begin{itemize}[nosep,leftmargin=*]
        \item \textbf{Activity:} Networking and reflection. \vspace{-1mm}
    \end{itemize}

    \textbf{3:30 PM – 4:30 PM: Group Activity: Design a Efficient Training Plan\vspace{-1mm}}
    \begin{itemize}[nosep,leftmargin=*]
        \item \textbf{Session:}
              \begin{itemize}[nosep,leftmargin=*]
                  \item \textbf{Task:} Small groups design a training plan for a hypothetical LLM project, incorporating efficiency strategies.
                  \item \textbf{Facilitator:} Provide templates and guide groups through the process.
              \end{itemize}
        \item \textbf{Takeaway:} Participants apply knowledge to real-world scenarios. \vspace{-1mm}
    \end{itemize}

    \textbf{4:30 PM – 5:00 PM: Closing Discussion \& Q\&A\vspace{-1mm}}
    \begin{itemize}[nosep,leftmargin=*]
        \item \textbf{Session:}
              \begin{itemize}[nosep,leftmargin=*]
                  \item \textbf{Facilitator:} Recap key takeaways and address remaining questions.
                  \item \textbf{Open Discussion:} Participants share insights, challenges, and next steps.
                  \item \textbf{Feedback:} Collect input for future workshops.
              \end{itemize}
        \item \textbf{Takeaway:} Participants leave with actionable strategies and a network of peers. \vspace{-1mm}
    \end{itemize}

    \textbf{Materials Needed\vspace{-1mm}}
    \begin{itemize}[nosep,leftmargin=*]
        \item Projector and screen for presentations.
        \item Printed handouts with key concepts, case studies, and templates. \vspace{-2.1mm}
    \end{itemize}
\end{tcolorbox}

\begin{tcolorbox}[colback=covenantwhite, colframe=covenantred, fonttitle=\bfseries, title=Prompt]
    \small Plan a moving day for a one-bedroom apartment with two helpers and one rented van. Give a realistic sequence of tasks.
\end{tcolorbox}

\begin{tcolorbox}[colback=white, colframe=covenantblk, fonttitle=\bfseries, title=Response]
    \small\setlength{\parskip}{0.5em}\setlength{\parindent}{0pt}
    To plan a moving day for a one-bedroom apartment with two helpers and one rented van, we'll break down the tasks into manageable chunks, ensuring a smooth and efficient process.

    First, start by preparing the night before the move. This includes packing non-essential items, such as seasonal decorations, off-season clothing, and books. Label and organize boxes by room to make unpacking easier. Also, prepare a "first night" box with essentials like toiletries, medications, and a change of clothes for each person.

    On moving day, begin by having breakfast and gathering all necessary supplies, including packing materials, tools, and the van keys. Assign tasks to each helper to maximize efficiency. One helper can focus on disassembling furniture, while the other assists with loading boxes and heavy items.

    Next, start loading the van, beginning with the heaviest and largest items, such as furniture and appliances. Load boxes in an organized manner, placing heavier boxes on the bottom and lighter ones on top. Make sure to secure everything with straps or rope to prevent shifting during transport.

    While loading the van, have one helper stay behind to pack any remaining items, such as dishes, linens, and cleaning supplies. Use bubble wrap, paper, or towels to protect fragile items.

    Once the van is loaded, conduct a final walk-through of the apartment to ensure everything is packed and nothing is left behind. Take photos of electronics and furniture before disassembling them to make reassembly easier.

    Before leaving the apartment, take final meter readings, and notify the landlord or property manager. Also, double-check that all doors and windows are locked, and the apartment is secure.

    Finally, arrive at the new apartment and begin unloading the van, starting with the largest and heaviest items. Assign tasks to helpers, such as assembling furniture and unpacking boxes. Take breaks as needed, and stay hydrated and energized throughout the day.

    Throughout the day, maintain open communication with helpers, and be flexible in case of unexpected delays or challenges. With a clear plan and efficient execution, the moving day should be a success.
\end{tcolorbox}

\subsection{Commonsense / Science}

\begin{tcolorbox}[colback=covenantwhite, colframe=covenantred, fonttitle=\bfseries, title=Prompt]
    \small Why is the sky blue during the day but red at sunset?
\end{tcolorbox}

\begin{tcolorbox}[colback=white, colframe=covenantblk, fonttitle=\bfseries, title=Response]
    \small\setlength{\parskip}{0.5em}\setlength{\parindent}{0pt}
    The reason the sky appears blue during the day and red at sunset is due to a phenomenon called \textbf{Rayleigh scattering}.

    {\normalsize\textbf{Here's how it works:}}

    \begin{enumerate}[nosep,leftmargin=*]
        \item \textbf{Sunlight is made up of all the colors of the visible spectrum} --- red, orange, yellow, green, blue, indigo, violet --- each with different wavelengths.
        \item \textbf{Rayleigh scattering} explains how molecules and small particles in the atmosphere scatter sunlight.
              \begin{itemize}[nosep,leftmargin=*]
                  \item Shorter wavelengths (like \textbf{blue and violet}) are scattered \textbf{much more} than longer wavelengths (like \textbf{red and orange}).
                  \item Our eyes are more sensitive to the color \textbf{blue} than to violet, so we perceive the sky as blue.
              \end{itemize}
        \item \textbf{During the day}, sunlight passes through the atmosphere and is scattered in all directions. Since blue light is scattered so much, we see it everywhere in the sky --- that's why the sky looks blue.
    \end{enumerate}

    {\normalsize\textbf{Why is the sky red at sunset?}}

    \begin{itemize}[nosep,leftmargin=*]
        \item \textbf{At sunset}, the sun is low on the horizon. Sunlight has to travel through \textbf{much more of the Earth's atmosphere} to reach your eyes.
        \item By the time sunlight reaches you at sunset, the \textbf{blue and green light has been scattered out of your line of sight}.
        \item The \textbf{longer, redder wavelengths} (red, orange, yellow) are less scattered and are able to reach you directly --- or after being reflected off particles --- giving the sky that warm, golden-to-red color.
    \end{itemize}

    {\normalsize\textbf{Bonus: Why are sunsets sometimes orange or pink?}}

    \begin{itemize}[nosep,leftmargin=*]
        \item The exact colors depend on atmospheric conditions.
              \begin{itemize}[nosep,leftmargin=*]
                  \item \textbf{Dust, pollution, or water droplets} can scatter longer wavelengths more, enhancing oranges and reds.
                  \item Cloud cover can reflect and amplify these colors, making sunsets dramatic.
              \end{itemize}
        \item The sun itself may appear larger or more orange due to atmospheric refraction and perspective.
    \end{itemize}

    So in short:\\
    \textbf{Blue sky = blue light scattered all around you.}\\
    \textbf{Red sunsets = blue light scattered away, leaving red light to reach your eyes.}

    It's a beautiful reminder of how physics gives us art in the sky.
\end{tcolorbox}

\begin{tcolorbox}[colback=covenantwhite, colframe=covenantred, fonttitle=\bfseries, title=Prompt]
    \small Why does a sealed bag of chips puff up on an airplane?
\end{tcolorbox}

\begin{tcolorbox}[colback=white, colframe=covenantblk, fonttitle=\bfseries, title=Response]
    \small\setlength{\parskip}{0.5em}\setlength{\parindent}{0pt}
    A sealed bag of chips puffs up on an airplane due to the decrease in air pressure at high altitudes. As the plane ascends, the air pressure outside the bag decreases, causing the air inside the bag to expand. Since the bag is sealed, the expanding air has nowhere to escape, resulting in the puffing up of the bag. This phenomenon is often referred to as "bag inflation" or "bag expansion." The expansion is a result of the difference in pressure between the inside and outside of the bag, and it's a common occurrence in sealed containers at high altitudes.
\end{tcolorbox}

\subsection{Instruction Following}

\begin{tcolorbox}[colback=covenantwhite, colframe=covenantred, fonttitle=\bfseries, title=Prompt]
    \small Give exactly 4 bullet points on the benefits and risks of open-source AI. Each bullet must be 12 to 18 words.
\end{tcolorbox}

\begin{tcolorbox}[colback=white, colframe=covenantblk, fonttitle=\bfseries, title=Response]
    \small\setlength{\parskip}{0.5em}\setlength{\parindent}{0pt}
    \begin{itemize}[nosep,leftmargin=*]
        \item Benefits: Open-source AI fosters innovation and collaboration, allowing diverse developers to contribute and improve models.
        \item Risks: The open nature can expose vulnerabilities, increasing the risk of malicious exploitation and data breaches.
        \item Benefits: Open-source AI promotes transparency and accountability, enabling users to inspect and trust the underlying algorithms.
        \item Risks: The lack of centralized control may lead to inconsistent or unreliable AI solutions, complicating adoption and standardization.
    \end{itemize}
\end{tcolorbox}

\begin{tcolorbox}[colback=covenantwhite, colframe=covenantred, fonttitle=\bfseries, title=Prompt]
    \small Return a JSON object with keys problem, assumptions, solution, and next\_steps for the task of reducing GPU training cost in a small lab.
\end{tcolorbox}

\begin{tcolorbox}[colback=white, colframe=covenantblk, fonttitle=\bfseries, title=Response]
    \setlength{\parskip}{0.5em}\setlength{\parindent}{0pt}
\begin{lstlisting}[style=covenantpython]
{
  "problem": "High GPU training costs in a small lab due to inefficient resource utilization and lack of optimization strategies.",
  "assumptions": [
    "The lab has limited GPU resources (e.g., a single GPU or a small cluster).",
    "Training jobs are often long-running and consume significant GPU memory.",
    "There is no existing automation for resource management or optimization.",
    "The team is familiar with basic machine learning concepts but lacks expertise in GPU-specific optimizations."
  ],
  "solution": "Implement GPU resource optimization strategies to reduce training costs, including job scheduling, memory management, and model quantization.",
  "next_steps": [
    "Conduct an audit of current GPU usage patterns to identify bottlenecks.",
    "Explore model quantization techniques to reduce memory footprint.",
    "Set up a job scheduling system to prioritize high-priority tasks.",
    "Monitor GPU utilization in real-time to adjust resource allocation dynamically."
  ]
}
\end{lstlisting}
\end{tcolorbox}

\subsection{Coding}

\begin{tcolorbox}[colback=covenantwhite, colframe=covenantred, fonttitle=\bfseries, title=Prompt]
    \small Write a Python function that implements topological sort and raises an error if the graph has a cycle. Explain your approach and its time complexity. Write smart and efficient code.
\end{tcolorbox}

\begin{tcolorbox}[colback=white, colframe=covenantblk, fonttitle=\bfseries, title=Response]
    \small\setlength{\parskip}{0.5em}\setlength{\parindent}{0pt}
    Here's a Python function implementing topological sort using Kahn's algorithm. This function raises a ValueError if the graph has a cycle.

\begin{lstlisting}[style=covenantpython]
from collections import deque

def topological_sort(graph):
    """
    Performs a topological sort on a directed acyclic graph (DAG).

    Args:
        graph (dict): A dictionary representing the graph, where each key is a node and its value is another dictionary of its neighbors.

    Returns:
        list: A list of nodes in topological order.

    Raises:
        ValueError: If the graph has a cycle.
    """
    in_degree = {node: 0 for node in graph}
    for node in graph:
        for neighbor in graph[node]:
            in_degree[neighbor] += 1

    queue = deque([node for node in graph if in_degree[node] == 0])
    sorted_nodes = []

    while queue:
        node = queue.popleft()
        sorted_nodes.append(node)
        for neighbor in graph[node]:
            in_degree[neighbor] -= 1
            if in_degree[neighbor] == 0:
                queue.append(neighbor)

    if len(sorted_nodes) != len(graph):
        raise ValueError("Graph has a cycle")

    return sorted_nodes
\end{lstlisting}

    \textbf{Approach:}

    1.  We first calculate the in-degree of each node in the graph.
    2.  We use a queue to keep track of nodes with an in-degree of 0, which are the starting points for our topological sort.
    3.  We iterate through the queue, adding nodes to the sorted list and decrementing the in-degree of their neighbors. If a neighbor's in-degree becomes 0, we add it to the queue.
    4.  If the length of the sorted list is not equal to the number of nodes in the graph, it means there's a cycle, and we raise a ValueError.

    \textbf{Time Complexity:}

    The time complexity of this algorithm is O(V + E), where V is the number of nodes and E is the number of edges in the graph. This is because we visit each node and edge once.

    \textbf{Example Usage:}
\begin{lstlisting}[language=Python]
graph = {
    'A': {'B', 'C'},
    'B': {'D'},
    'C': {'D'},
    'D': {}
}
sorted_nodes = topological_sort(graph)
print(sorted_nodes)  # Output: ['A', 'C', 'B', 'D']
\end{lstlisting}
     \vspace{-2mm}This code assumes that the graph is represented as a dictionary where each key is a node and its value is another dictionary of its neighbors. The function returns a list of nodes in topological order. If the graph has a cycle, it raises a ValueError.
\end{tcolorbox}

\subsection{LaTeX Conversion}

\begin{tcolorbox}[colback=covenantwhite, colframe=covenantred, fonttitle=\bfseries, title=Prompt]
    \small Convert this to LaTeX: The matrix A = [[1, 2], [3, 4]] has determinant -2 and inverse (1 / -2) [[4, -2], [-3, 1]].
\end{tcolorbox}

\begin{tcolorbox}[colback=white, colframe=covenantblk, fonttitle=\bfseries, title=Response]
    \small\setlength{\parskip}{0.5em}\setlength{\parindent}{0pt}
\begin{Verbatim}[fontsize=\small]
Here is the LaTeX representation of the given matrix and its properties:

\[
\boxed{
\begin{aligned}
A &= \begin{pmatrix} 1 & 2 \\ 3 & 4 \end{pmatrix}, \\
\det(A) &= -2, \\
A^{-1} &= \frac{1}{-2} \begin{pmatrix} 4 & -2 \\ -3 & 1 \end{pmatrix}.
\end{aligned}
}
\]
\end{Verbatim}
\end{tcolorbox}

\begin{tcolorbox}[colback=covenantwhite, colframe=covenantred, fonttitle=\bfseries, title=Prompt]
    \small A function f is called increasing if f(s) <= f(t) holds for every s <= t. Convert this paragraph to LaTeX.
\end{tcolorbox}

\begin{tcolorbox}[colback=white, colframe=covenantblk, fonttitle=\bfseries, title=Response]
    \small\setlength{\parskip}{0.5em}\setlength{\parindent}{0pt}
\begin{Verbatim}[fontsize=\small]
A function $ f $ is called increasing if $ f(s) \leq f(t) $
holds for every $ s \leq t $.
\end{Verbatim}
\end{tcolorbox}

\subsection{Creative Writing}

\begin{tcolorbox}[colback=covenantwhite, colframe=covenantred, fonttitle=\bfseries, title=Prompt]
    \small Write the opening scene of a mystery novel in which the detective is also the prime suspect.
\end{tcolorbox}
\begin{tcolorbox}[colback=white, colframe=covenantblk, fonttitle=\bfseries, title=Response]
    \small\setlength{\parskip}{0.5em}\setlength{\parindent}{0pt}
    The rain-soaked streets of Ravenswood were always at their most unforgiving, but tonight they seemed to be closing in on Detective James Winter like a shroud. He stood outside the grand estate of Henry Blackstone, the wealthy businessman whose murder had left the town reeling. The police tape fluttered in the wind, and the faint scent of cigar smoke still lingered in the air, a grim reminder of the events that had unfolded just hours before.

    As James watched, a team of forensic experts carefully processed the scene, their flashlights casting eerie shadows on the walls. He felt a familiar knot in his stomach, one he'd grown accustomed to over the years. This was his territory, his domain – but tonight, he was the one under scrutiny. The victim's eyes had been frozen in a permanent stare, a look that seemed to be accusing him, even as the police began to whisper about the detective's troubled past.

    James's mind was a jumble of memories, each one more disturbing than the last. He'd been at the Blackstone estate just hours before the murder, arguing with the victim over a case that had gone cold. The tension had been palpable, the air thick with unspoken threats. And now, as he stood there, he couldn't shake the feeling that he was being watched, that the eyes of the town were upon him.

    A voice cut through the night, sharp and insistent. "Detective Winter? We need to talk." James turned to see a young officer, her eyes narrowed with suspicion. He forced a smile, but his heart was racing. He knew that look – it was the same one he'd seen on the faces of his colleagues, the same one that said, "We know you're hiding something."

    As James followed the officer to a nearby car, he couldn't help but wonder: had he finally crossed the line? Had his obsession with the Blackstone case led him to commit the very crime he was sworn to prevent? The rain poured down, washing away the evidence, but not the doubts that had been creeping into his mind. Tonight, James Winter was the prime suspect, and he was running out of time to prove his innocence.
\end{tcolorbox}

\begin{tcolorbox}[colback=covenantwhite, colframe=covenantred, fonttitle=\bfseries, title=Prompt]
    \small Write a fairy tale about a clock that refuses to tell time.
\end{tcolorbox}

\begin{tcolorbox}[colback=white, colframe=covenantblk, fonttitle=\bfseries, title=Response]
    \small\setlength{\parskip}{0.5em}\setlength{\parindent}{0pt}
    In a small village nestled between two great mountains, there stood an ancient clock tower that had been the village's pride for generations. The clock, named Chronos, was said to have been crafted by the finest clockmakers of the land, and its face shone like polished silver in the sunlight. For years, Chronos had kept perfect time, its rhythmic tick-tock a comforting heartbeat for the villagers.

    But one day, something strange happened. Chronos began to falter, its gears grinding to a halt as the clock face froze at 3:47. The villagers were perplexed, but they soon realized that Chronos was not broken – it was simply refusing to tell time.

    At first, the villagers were frustrated, relying on their own pocket watches and sundials to keep track of the hours. But as the days passed, they began to notice something peculiar. Without the constraints of time, the villagers found themselves living in the moment. They would gather in the town square, sharing stories and laughter without worrying about the hour. Children would play all day, their imaginations running wild without the pressure of a ticking clock.

    The village elder, a wise and kind woman named Aria, saw the beauty in Chronos's refusal to tell time. She realized that the clock was teaching the villagers a valuable lesson: that time is not a rigid framework, but a fluid and ever-changing force that should be savored, not rushed.

    As the weeks went by, the villagers grew to appreciate Chronos's unique gift. They would gather around the clock tower, listening to the gentle hum of its gears, and feel the weight of time lifting off their shoulders. The clock's refusal to tell time had become a reminder to live in the present, to cherish every moment, and to never take a single second for granted.

    And so, Chronos continued to stand tall, its face frozen at 3:47, a symbol of the village's newfound appreciation for the beauty of timelessness. The villagers would often say that Chronos was not broken, but rather, it was perfect – a reminder that sometimes, the best way to tell time is to forget it altogether.
\end{tcolorbox}

\subsection{Multi-audience Explanation}

\begin{tcolorbox}[colback=covenantwhite, colframe=covenantred, fonttitle=\bfseries, title=Prompt]
    \small Explain overfitting to a middle school student, a new ML student, and a research scientist.
\end{tcolorbox}

\begin{tcolorbox}[colback=white, colframe=covenantblk, fonttitle=\bfseries, title=Response]
    \small\setlength{\parskip}{0.5em}\setlength{\parindent}{0pt}
    Here's an explanation of overfitting for different audiences:

    \textbf{Middle School Student:}

    Imagine you're trying to guess someone's favorite ice cream flavor based on their favorite color. You collect data on people's favorite colors and their favorite ice cream flavors. After analyzing the data, you come up with a rule: if someone's favorite color is blue, their favorite ice cream flavor is chocolate. However, this rule doesn't work for everyone. Some people with blue favorite colors might like vanilla, and some people with other favorite colors might like chocolate. Overfitting is when your rule is too specific and only works for the data you used to create it, but not for new data. It's like trying to fit a square peg into a round hole.

    \textbf{New ML Student:}

    Overfitting occurs when a model is too complex and learns the noise in the training data, rather than the underlying patterns. This results in a model that performs well on the training data but poorly on new, unseen data. Think of it like memorizing the answers to a practice test instead of learning the material. The model is essentially "memorizing" the training data, including the noise and random fluctuations, rather than learning the generalizable patterns. To avoid overfitting, we use techniques like regularization, cross-validation, and early stopping to ensure the model is generalizing well to new data.

    \textbf{Research Scientist:}

    Overfitting is a fundamental problem in machine learning where a model becomes overly specialized to the training data, capturing noise and random fluctuations rather than the underlying patterns. This is often characterized by a significant gap between the model's performance on the training data and its performance on unseen data. Overfitting can be attributed to various factors, including model complexity, data quality, and training procedures. To mitigate overfitting, researchers employ a range of techniques, including regularization, early stopping, and ensemble methods, to balance model complexity and generalization performance. The goal is to develop models that can effectively generalize to new, unseen data, while minimizing the risk of overfitting.
\end{tcolorbox}

\end{document}